\documentclass[10pt, conference, compsocconf]{IEEEtran}
\usepackage[T1]{fontenc}
\usepackage{float}
\usepackage{amstext}
\usepackage{verbatim}
\usepackage{textcomp} 

\usepackage[cmex10]{amsmath}
\usepackage{algorithm}
\usepackage{algpseudocode}
\floatstyle{ruled}
\newfloat{algorithm}{tbp}{loa}
\floatname{algorithm}{Algorithm}

\algnotext{EndFor}
\algnotext{EndIf}
\usepackage[numbers]{natbib}

\usepackage{graphicx}

\usepackage{mdwlist}
\usepackage{paralist}

\makeatletter
\renewcommand\section{\@startsection{section}{1}{\z@}%
                       {5\p@ \@plus 2\p@ \@minus 2\p@}%
                       {5\p@ \@plus 2\p@ \@minus 2\p@}%
                       {\large \bfseries \raggedright}}
\renewcommand\subsection{\@startsection{subsection}{2}{\z@}
{5\p@ \@plus 4\p@ \@minus 4\p@}%
{3\p@ \@plus 4\p@ \@minus 4\p@}%
{\normalsize \bfseries \raggedright}}
\makeatother
\usepackage{url}
\usepackage{url}
\makeatother

{\footnotesize \bibliographystyle{abbrv}}
\usepackage{array}
\usepackage{mdwmath}
\usepackage{mdwtab}
\usepackage{eqparbox}
\usepackage[tight,footnotesize]{subfigure}

\begin{document}
%
% paper title
% can use linebreaks \\ within to get better formatting as desired
\title{Distributed Wear levelling of Flash Memories}

% author names and affiliations
% use a multiple column layout for up to two different
% affiliations

\author{%
% author names are typeset in 11pt, which is the default size in the author block
{Srimugunthan{\small $~^{1}$}, K. Gopinath {\small $~^{2}$} }%
% add some space between author names and affils
\vspace{1.6mm}\\
\fontsize{10}{10}\selectfont\itshape
 Computer Science and Automaton Department, Indian Institute of Science,\\
 Bangalore, India\\
\fontsize{9}{9}\selectfont\ttfamily\upshape
$~^{1}$srimugunth@csa.iisc.ernet.in
$~^{2}$gopi@csa.iisc.ernet.in\\

}

% make the title area
\maketitle

\begin{abstract}
\textit{
For large scale distributed storage systems, flash memories are an excellent choice because flash memories consume less power,
take lesser floor space for a target throughput and provide faster access to data. 
 In a traditional distributed filesystem,  even distribution is required to ensure load-balancing, balanced 
space utilisation and failure tolerance. In the presence of flash memories, in addition, we should also ensure that the number of writes
 to these different flash storage nodes  are evenly distributed, to ensure even wear of flash storage nodes, so that unpredictable failures 
of storage nodes are avoided. This requires that we distribute  updates and do garbage collection,  across the flash storage nodes. We have motivated the
 distributed wearlevelling problem considering the replica placement algorithm for HDFS. Viewing the wearlevelling across flash storage nodes as a distributed 
co-ordination problem,  we present an alternate design, to reduce the message communication cost across participating nodes. We demonstrate the effectiveness of
 our design through simulation. 
}

\end{abstract}

\begin{comment}
\begin{IEEEkeywords}
Mass Storage; Storage Management; Distributed Memories; FileSystems Management; Distributed file systems;

\end{IEEEkeywords}
\end{comment}

% For peer review papers, you can put extra information on the cover
% page as needed:
% \ifCLASSOPTIONpeerreview
% \begin{center} \bfseries EDICS Category: 3-BBND \end{center}
% \fi
%
% For peerreview papers, this IEEEtran command inserts a page break and
% creates the second title. It will be ignored for other modes.
%\IEEEpeerreviewmaketitle

\section{Introduction}
% no \IEEEPARstart

	Flash memory is different from a hard disk storage because of the following peculiarities.
	Every write can happen only on an erased block. The granularity of erase(in terms of blocks)
 is much bigger than the write or read(in terms of pages). The amount of erase that a block can sustain
 before becoming bad is in hundreds of thousands. To take care of flash specific eccentricities,
 wear leveling is done that makes sure that number of erases that happen to blocks increase uniformly.

Flash memories are currently used in Portable media players, laptops. As the cost of flash memories 
come down, they will be  used more ubiquitously. For large scale storage systems, flash memories
 are an excellent choice. A previous work \cite{Gordon09} evaluated the usage of flash memories
 in high performance clusters and have showed the advantage of flash with respect to cost and performance/watt. 

      When flash memories are used in distributed storage system, we also need to address the uneven 
wear-out that can happen between different flash storage nodes.
	
      The data distribution with traditional hard disks are  concerned with 
 \textit{balanced space utilisation}, ie ratio of the amount of data stored to available 
storage on a disk  should be balanced  across all disks in the cluster    and \textit{load-balancing} ie number
 of references or accesses to a disk  is balanced across all disks in the cluster  to prevent network hot-spots. 
      But they don't consider flash specific wear leveling.

 For enterprise workloads, the lifetime of the SSD drive  which is greater 
than 10GB,  is predicted in  \cite{ssdenterprise09} to be greater than 5 years. 
Table \ref{tab:WearOut} shows the theoretical wear out times for different sizes of flash, with different 
maximum erase cycles and with different writing speeds. For larger size flash drives, because of 
longer lifetimes, wearlevelling across flash drives is not a significant concern. But for a system made of 
 smaller size flash memory chips, wearlevelling across flash chips is important.

\begin{table*}[]
\centering

    \caption{Time for wear out for different flash sizes}    
    \label{tab:WearOut}

    \begin{small}
    \begin{tabular}{|l|l|l|l||l|l|l|}
    \hline
    {\bfseries Size of} & \multicolumn{3} {c|} {\bfseries Max Wear = 10,000 erase cycles} &  \multicolumn{3} {c|}  {\bfseries Max Wear = 100,000 erase cycles}\\
    \cline{2-4} \cline{5-7}
    {\bfseries Storage nodes} & {\bfseries  40MB/s}         & {\bfseries 80MB/s}     & {\bfseries 100MB/s}  & {\bfseries  40MB/s}         & {\bfseries 80MB/s}     & {\bfseries 100MB/s}          \\
    \hline
                  
    \hline
    1GB        & 2.96 days 	&	1.48 days	& 1.18 days  & 29.63 days & 14.81 days & 11.85 days    \\

    \hline
    2GB        &	5.93 days 	&	2.96 days 	&	2.37 days & 59.62 days & 29.63 days & 23.70 days \\
    \hline

    5GB        &	14.81 days	&	7.41 days 	&	5.93 days & 148.15 days & 74.07 days & 59.26 days\\
    \hline

    10GB        &	29.63 days 	&	14.81 days 	&	11.85 days & 296.30 days & 148.15 days & 118.52 days\\
    \hline

    \hline
    \end{tabular}
    \end{small} 
\end{table*}

 The maximum capacity available in the market, as of this writing is 16GB for flash memory chips
 and a terabyte for SSD drives.
Designs are evolved to use flash memories in server environment. 
There are three usage scenarios for flash memories proposed for servers in \cite{Roberts09}. 
Flash memory chips can be used as extended system memory, as a PCI express card acting as disk 
cache or as SSD drives  completely  substituting the disks.  An interesting scenario proposed 
in \cite{Gordon09} is to have a sub-cluster with nodes made of flash storage like a specialised 
co-processor meant for fast data processing. This means that there will be a bigger cluster made of
 disks and there is a smaller compute sub cluster made of small sized flash memories. Since flash memory
 is good for IOPS per dollar and disks are good for gigabyte per dollar, these kind of tradeoff designs 
can prove advantageous. For the  kind of scenario, where the storage nodes in the cluster are made of small size flash memory
 chips(less than 10GB), making  the wear-out across flash storage nodes even is important to avoid unpredictable failures. 
Even for large sized SSD drives  distributed wear leveling will result in longer lifetime of
 the individual drives.

The organisation of the paper is as follows. Section 2 gives the background with respect to wearlevelling and distributed file systems. Section 3 describes the motivation for distributed wear levelling. Section 4 describes a design for distributed wear levelling, Section 5 gives the simulation results. Section 6 discusses some related work.  Section 7  discusses some possible future work and concludes.
% You must have at least 2 lines in the paragraph with the drop letter
% (should never be an issue)

\section{Background}

\subsection{Wearlevelling in Flash memories}
Wearleveling algorithms, in addition to balancing the writes across the blocks in the nand device, have several other goals like
\begin{itemize}
 \item minimising the extra table space taken by the flash translation layer
 \item minimising the garbage collection overhead
 \item minimising the data migration within the flash device.
\end{itemize}
 
These objectives are influenced by the algorithm that does the address translation, the policies that select which block to write next, the policies that decide when and how to do garbage collection, in policies that decide what kind of data is written to a particular block etc.

 There are excellent papers published, which address the problem of wear leveling with in a flash. 

\subsection{Distributed File systems: Brief survey}
A distributed filesystem provides a single namespace for accessing storage 
disks that is dispersed across a network. A distributed filesystem mounts 
all the dispersed storage under a single mount point.
  Some of the requirements of distributed filesystems are incremental scalability, failure tolerance, location transparency etc.

Examples of  distributed filesystems are Coda, Lustre, Hadoop distributed file system (HDFS), Google file system (GFS) and Ceph.\newline
 
\textbf{Coda:} Coda is a relatively older distributed filesystem with a client-server architecture. 
The actual storage  is present in the file servers and clients make their I/O request to the file/storage servers.
 A file is mapped on to a storage server. To tolerate failures of storage nodes, Coda does replication of data.
 The unit of replication in coda is a `volume', which is a subtree of the namespace. The main feature of Coda is
 aggressive  caching by the client nodes, which helps  the clients to operate, even if they are disconnected from
 the network. Modifications to client-cached files are intimated to file servers through callback messages. Coda provides
 transactional semantics in the case of concurrent read-write scenarios.  Perfect consistency guarantees any read of a file,
 gives the most recently written data by any client. This consistency requirement is relaxed in many filesystems, 
because studies in distributed filesystems have shown that concurrent write-sharing is rare enough \cite{measuredistfs}.
In Coda, a client which does a file open sees  modifications that was done by a some client's last file close.

Recent distributed filesystems are designed for scale supporting tens of petabytes of storage, and hundreds of 
gigabytes-per-second of aggregate bandwidth. Another important requirement is that the design has to cope with 
failure-as-norm scenarios where  everyday one or two  storage nodes fail per thousand nodes\cite{HDFSpaper}. 
A distributed filesystem workload comprises a lot of concurrent metadata operations, than concurrent data operations.
 For this reason, recent distributed file systems have the architecture of clients, storage servers and a few management
 servers. The management servers are used only for  filesystem metadata operations like pathname lookup, file creation etc. 
Having separate dedicated servers for metadata operations, helps avoid bottlenecks that arises from locking for concurrent
 operations. The metadata operations can be more quickly done than data operations and they comprise more than 50\% of filesystem operations.
It is also easy for doing filesystem consistency checks if we have separate servers for filesystem metadata operations.
 Hence having separate servers for filesystem metadata operations is a `compelling design' for a distributed filesystem. \newline

\textbf{Lustre:}
Lustre follows the architecture of clients, metadata servers and object storage targets.
Lustre assumes an object storage model for individual storage servers. In the case of object based storage devices, 
data can be written  as variable size objects instead of fixed sized sectors. 
Lustre requires individual disks to support object interface through hardware or by a separate software driver that 
exposes a object based interface.  Lustre implements its own network stack. The Lustre network stack
 is optimised for high performance network transfers like RDMA and also supports heterogeneous interconnects.
 The metadata server stores the information corresponding to the location of objects for a file.
 At file open time, the clients hash the filename to obtain the responsible
 metadata server. The metadata server intimates the client, the storage servers where the file is mapped to. 
Subsequently the client  can do  I/O directly  from  the storage server without further involvement of the metadata server. 
Lustre allows multiple concurrent readers or writers of a file. Lustre  follows the distributed locking management, with 
the inode metadata related locks handled only by the metadata server and  each storage server  handles the locking of  
objects it stores. So there are different types of locks and depending on the specific kind of operation a client need to 
perform, it fetches the  particular kind of lock.

\textbf{HDFS and GFS:}
 Google file system and HDFS follow similar designs and are mainly targeted for \textit{write-once and read-many} streaming
 workload which are typical of map reduce applications. The files are stored as relatively larger blocks(64MB-256MB) on the
 storage servers. The original HDFS and GFS used a single masternode design with all the metadata such as file to blocks mappings
 and  block to storage server mappings, stored by a single central master node. However, recent changes in design use multiple master
 nodes with each of them responsible for a subtree of the global filesystem namespace.
  
  In HDFS, the masternode is called the namenode. Namenode stores all the metadata information of the distributed filesystem and
 storage servers are called datanodes. Namenode also maintains a journal called `edit log' which stores all the transactions of the 
 filesystem. The clients and the datanodes, query and interact with the namenode for all operations. There is also a secondary namenode
 used for recovery when namenode fails. The datanode can host any local disk filesystem. The datanodes store the HDFS blocks as a file on the
local  filesystem. Every HDFS block is stored along with checksum, which is stored as another local file.   The filesystem read/write/append
 operations are generated by applications running at clients.  HDFS does not support file re-writes. Also two clients in HDFS cannot write on
 the same file simultaneously. 

%After all the target nodes are selected, nodes are organised as a pipeline, in the order of closeness to the first replica. Data is forwarded to nodes in this order. 

\begin{comment}
%\begin{figure*}
%\begin{center}
% \includegraphics[scale = 0.3]{./hdfs_rwflow.eps}
 % simpdsol.eps: 0x0 pixel, 300dpi, 0.00x0.00 cm, bb=14 14 267 171
%\caption{ hdfs }
%\label{fig:hdfs read write}
%\end{center}
%\end{figure*}
\end{comment}

The masternode detects failure of storage server, by  heart beat messages with a typical periodic interval of three seconds. 

  GFS is similar to HDFS in many aspects. Clients in GFS rely on the buffer caching of the operating system and don't explicitly cache 
file data. GFS and HDFS do their replica placement to balance the load and space utilisation across storage servers and to handle 
correlated failures. Because nodes that are connected to common power supply or a common router, can become unreacheable simultaneously, 
they are treated as nodes belonging to a common failure domain. The replica placement algorithm should spread the data across failure domains. 
 The HDFS replica placement is discussed in more detail in the next section.

In Google file system \cite{GFS03} data of new replicas are placed on  storage servers with below average disk space utilisation.
 Equalising space utilisation is important to equalise the future load on the servers and to not unfairly concentrate data on one disk,
 so that, failure of one disk doesn't cause loss of data. The replica placement algorithm makes sure that replicas are spread  
across machines to fully utilize each machine's bandwidth and spread replicas across racks to protect against correlated failures.
 It also does rebalancing, where data is migrated between storage servers for equalising space utilisation and load balancing.\newline

\textbf{Ceph:}
Ceph \cite{Ceph06} is a recent distributed object based filesystem, that is designed for general purpose I/O workload. 
The design of Ceph distributed filesystem  assumes the presence of  tens of metadata servers, object based storage servers and 
hundreds of thousands of clients. The central idea in the design of the Ceph  is the use of pseudo random CRUSH hashing function
 to map file objects to storage servers. The CRUSH hash  function \cite{CRUSH06} fulfills a chief requirement, which is that the 
hash function mapping  changes minimally when the range of the hash function changes. This means that  when the cluster configuration 
changes due to failure of storage nodes or due to addition of new storage, it changes the hash mapping minimally. This is important because changes in hash mapping
induces movement of  objects between the storage nodes. Minimising this data movement is the goal of a  data distribution function.
 In addition, Ceph's CRUSH mapping achieves statistical balancing by randomly distributing the data and also distributes the replicas across failure domains.

    Because of the use of CRUSH mapping function, the location of objects is well known and the clients don't 
have to contact the metadata server for this information. 

    The use of hashing function minimises the amount of metadata of a single object.  This makes possible the \textit{dynamic subtree partitioning}  feature of Ceph. 
In HDFS a single management server is statically responsible for a subtree of the  distributed filesystem namespace. 
In Ceph, the responsible management server for the subtree of the distributed filesystem namespace changes dynamically.
 This is a desirable feature, because sometimes  when many clients read the same file, the responsible management server can become overloaded.
 During heavy load on a management server, the metadata information is copied to other management servers. This avoids hot-spots on any one particular management server. \newline

\ifdefined\THESIS

\begin{figure}
\centering
\subfloat[]{\includegraphics[width=3.4in]{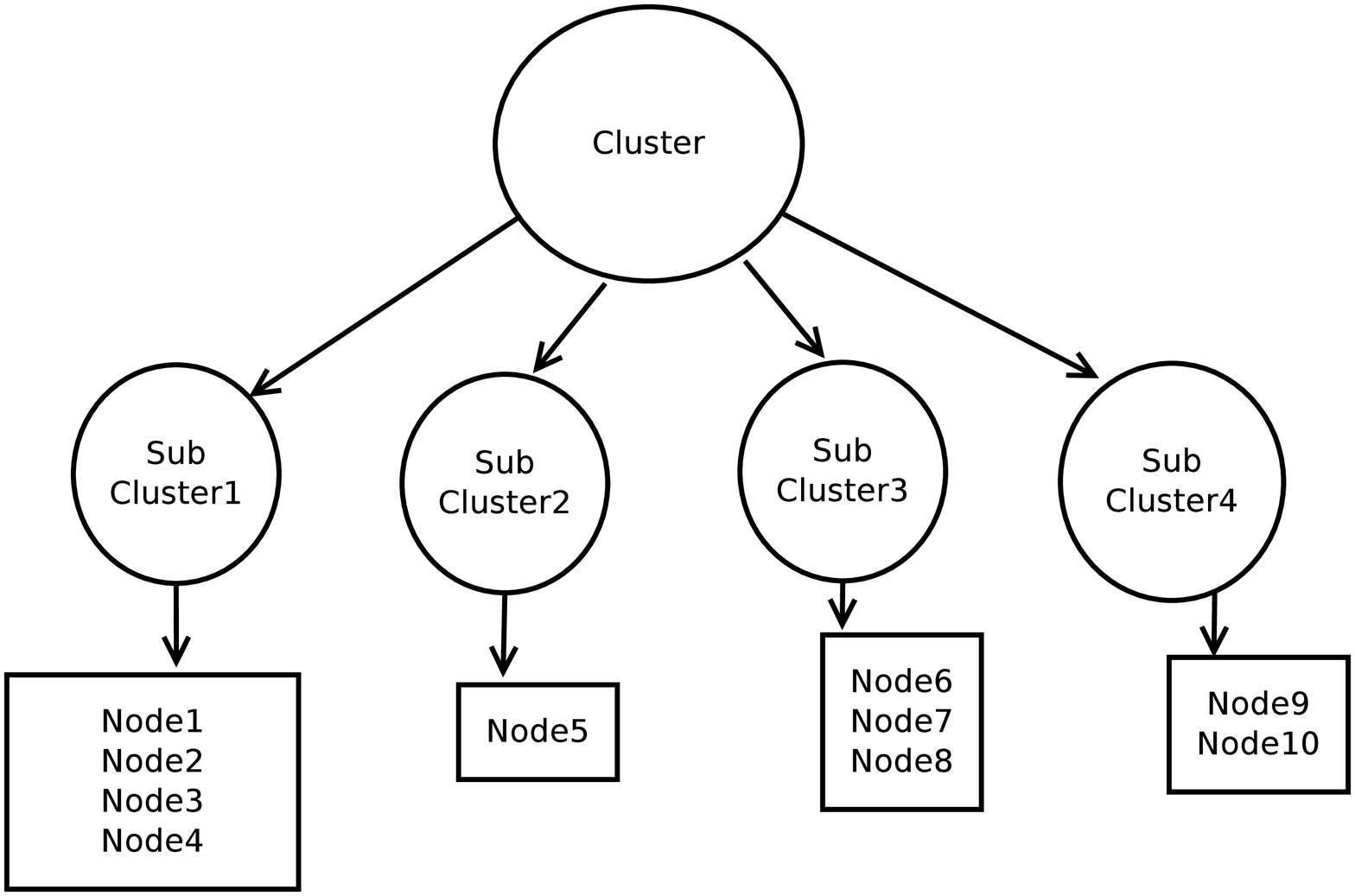}}
\subfloat[]{\includegraphics[width=3.4in]{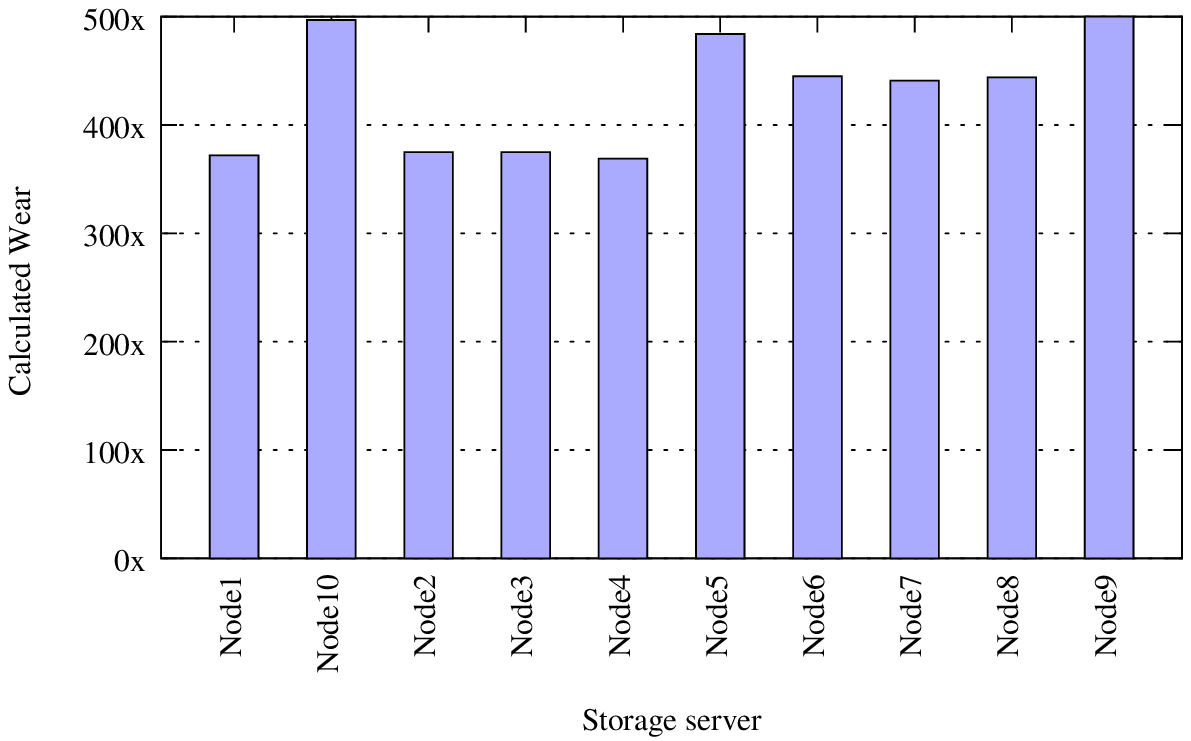}}
\caption{  (a)An example of a lop-sided Cluster Topology, (b) Calculated wear of the nodes in this topology.}
\label{fig:topolWear}
\end{figure}

\begin{comment}
\begin{figure}[htb]
  \vspace{9pt}

  \centerline{\hbox{ \hspace{0.0in} 
    \epsfxsize=3.4in
    \epsffile{./figures/cluster.eps}
    \hspace{0.25in}
    \epsfxsize=3.4in
    \epsffile{./figures/weargrph.eps}
    }
  }

  \caption{ This Figure contains two panels labelled (a)Cluster Topology, (b)Calculated wear.}
  \label{fig:topolWear}

\end{figure}
\end{comment}

\else

\begin{figure*}

\begin{minipage}[b]{0.5\columnwidth}
 \centering
\end{minipage}

\hspace{1.8cm}
\begin{minipage}[b]{0.5\columnwidth}

\includegraphics[width=1.8in,height=1.5in]{./figures/cluster.eps}
\caption{Cluster Topology}
\label{fig:cluster}
\end{minipage}
\hspace{0.2cm}
\begin{minipage}[b]{0.5\columnwidth}
 \centering
\includegraphics[width=1.5in,height=1.5in]{./figures/weargrph.eps}
\caption{Calculated wear}
\label{fig:calcwear}
\end{minipage}
\hspace{0.2cm}
\begin{minipage}[b]{0.5\columnwidth}
 \centering
\includegraphics[width=2in,height=1.5in]{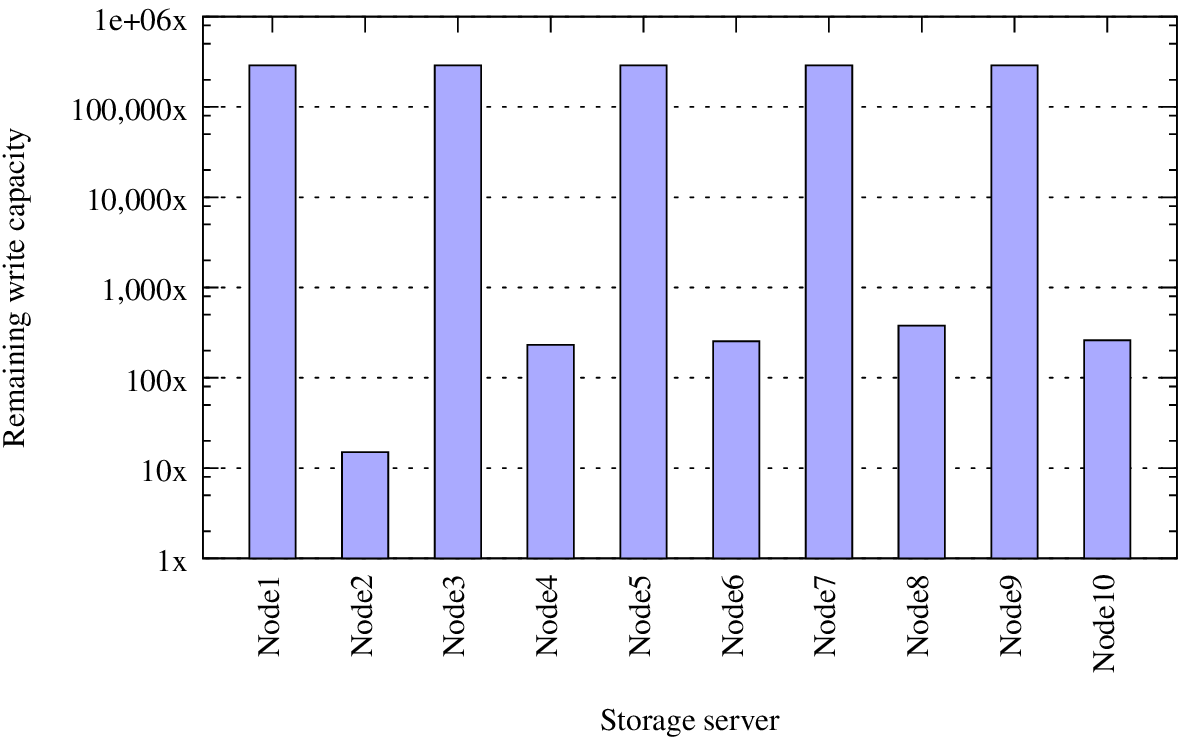}
\caption{Remaining write capacity}
\label{fig:remwrcap}
\end{minipage}
\end{figure*}
   
\fi

 Almost all distributed filesystems uses remote procedure calls for inter-communication messages between clients, management servers and storage servers. 
The minimum size of  a RPC communication message is atleast 64 bytes. The number of messages per second is limited by the kind of network interconnect in
 the distributed filesystem. Reducing the number of messages is important because, frequent communication messages  affects the linear scaling of I/O bandwidth
 and performance of a distributed filesystem and also hinder the responsiveness at the client end.
\begin{comment}
%In a heterogeneous storage system, balancing both load and space utilisation requires tradeoffs. In GPFS\cite{GPFS02}, the administrator should manually configure whether the data distribution algorithm should minimise load or space utilisation. 
 \end{comment}
  
 To summarise, the data distribution problem  in distributed file systems  when hard disks are used in the storage system is dominated by the following considerations
\begin{itemize}
\item  storing the replicas in different failure domains(not  in the same disk, or  in the same shelf with a common power supply)
\item  number of references or accesses   is balanced across all disks in the cluster  to prevent network hot-spots (ie \textit{balanced load})
\item   ratio of the amount of data stored to available storage on a disk  should be balanced  across all disks in the cluster (ie \textit{balanced space utilisation}). 
\item  optimal data movement due to addition or removal of storage
\end{itemize}
We view the equalising the wear across storage nodes, as a different requirement to the data distribution and the next section shows why this is so.

\subsection{Motivation for distributed wearlevelling}
%In the next section, we give the motivation for distributed wearleveling.
%\vspace{10 mm}

  For replica placement algorithms like Ceph, a object is mapped by a hashing function to  some storage server.
 If a particular file or object is over written  multiple times, wear-out across storage nodes will be uneven.

 A storage system is made of multiple racks, enclosures etc. A sub-cluster of nodes can be group of  nodes belonging the same failure
 domain like nodes within a rack or enclosure.  HDFS follows a replica placement algorithm which tries to spread the replicas
 across failure domains.  In addition, the HDFS replica placement takes in to account, the load and space utilisation of storage servers.

\begin{comment}
    %When re-replicating a block, if the number of existing replicas is one, place the second one on a different rack. When the number of existing replicas is two, it the replicas are on the same rack, place the third one on a different rack; otherwise place the third one on a different node at the same rack of the first replica. When the number of available replicas is more than two, place the replicas randomly.
%For reading, the namenode checks if the client's computer is located in the cluster. If yes, block locations are returned to the client in the order of its closeness to the reader. The block is read from datanodes in this preference order. 
\end{comment}

%newline
%\subsubsection{Topology based placement}
In HDFS, when a client writes a new block, it contacts the management server or the namenode for the locations of the storage servers which 
can be targeted for writing. The management server uses the following replica placement algorithm to decide a storage server for each replica 
of the block.

 In the following, the `local node' refers to the node that generates the write.  The `local rack' refers to the  rack that houses the local node.
The `remote rack' refers to any rack other than the local rack. The `remote node' refers to the node different from the local node.

  The first replica is placed on the local node. This is done so that future reads can be faster. 
The second replica is placed on a remote rack, and the third replica 
is placed on a remote node in the local rack. The rest of the replicas are placed randomly on some storage server.
 But the placement of nodes always follows the restriction that no more than one replica is placed at one node, and no more 
than 2 replicas is placed in the same rack. The above restriction is relaxed when the number of replicas is more than  twice
 the number of racks. After the selection of the replica target, it is checked if the target is good enough in terms of current
 load and space utilisation.  If the particular storage node is overloaded or if the remaining space is too low, then the replica
 placement algorithm again looks for another target node.

We simulated the replica placement algorithm and for our simulation, we assumed a workload with a  lot of small sized files created and deleted repeatedly.
 We assumed that all the storage nodes are of the same size.

  We calculated the \textit{current wear} of a particular storage node as ``((Total amount of bytes written so far))/(Storage node size))''. 
% We calculated the current wear of a particular storage node as (Number of block-writes to the storage node so far)/(Number of blocks in storage node). 

 Because of the topology based placement, it  happens that for certain lopsided cluster topologies data
 distribution can result in uneven wear-out across storage nodes. For example, the topology in Figure \ref{fig:cluster} , assumes that
 the cluster is made of 4 sub-clusters. A sub-cluster can be a set of nodes within a rack or shelf.
 Each sub-cluster has different number of nodes in it.  The  Figure \ref{fig:calcwear}  shows the unevenness of the wear at the end of simulation. 

    Even for a linear topology with storage nodes of same capacities, uneven wear-out can result if the storage nodes 
have different flash endurance limits.  For MLC nand chips the maximum endurance cycle is in tens of thousands, but for SLC the maximum 
endurance cycle limit is in hundreds of thousands. Between different flash vendors, endurance cycle limit varies. For our 
simulation we assumed half of the storage nodes to have maximum  endurance cycles as 5000 and the other half to have a maximum endurance cycles as 500.
We  used a metric `\textit{remaining write capacity}'  to denote the amount of writes that can still be done on a storage node, The `remaining write capacity' is 
equal to ((Amount of data written so far)/Block size) - ((Total blocks in storage node) * Maximum endurance cycles). The plot in Figure \ref{fig:remwrcap} shows
 the remaining write capacity of storage nodes at the end of simulation.
It shows that the  remaining write capacity of Node2 is the lowest  This means that  Node2 will fail early compared to other storage nodes.

 To overcome the uneven wear out we have to limit the amount of writes to a flash storage node depending on the
 remaining write capacity. In the next section we give a design that can be used for budgeting the writes to the storage servers.

\begin{comment}
 %We modified the replica placment algorithm  by attaching  a weight to each node in the topology tree. The weights are added based on the aggregate endurance cycles a sub-cluster or a flash storage node can take. When the replica placement algorithm selects one of the nodes or subcluster randomly, it respected the weights attached. 

%With the modified replica placement algorithm, we were able to get even wear across the storage nodes as shown in the plot.
\end{comment}

\ifdefined\THESIS

\begin{figure}[htb]
  \vspace{9pt}

  {\hbox{ \hspace{1.5in} 
    \epsfxsize=3.4in
    \epsffile{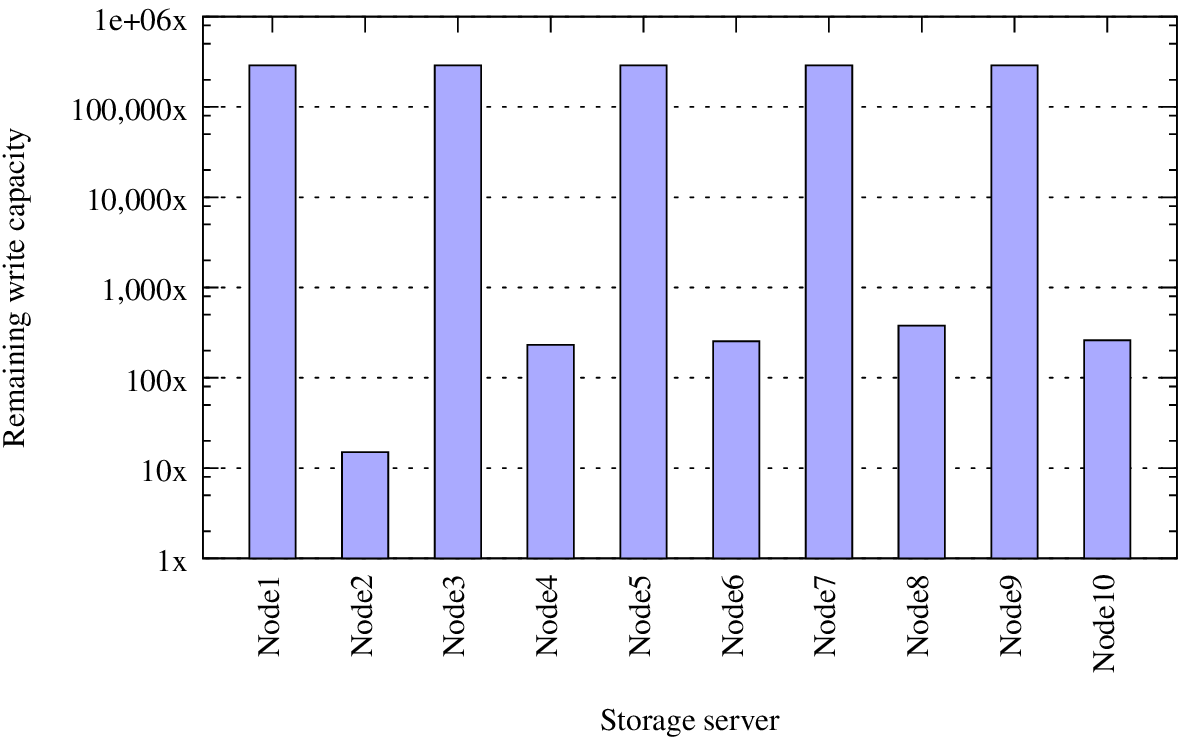}
    \hspace{0.25in}
    
    }
  }

  \caption{ Remaining write capacity. Node2 has the lowest remaining capacity compared to other node.}
  \label{fig:remwrcap}

\end{figure}

\FloatBarrier
\fi

\section{A design for Distributed Wearlevelling}

\begin{figure*}
%\begin{minipage}[b]{0.5\linewidth}
\centering
\includegraphics[scale=0.25]{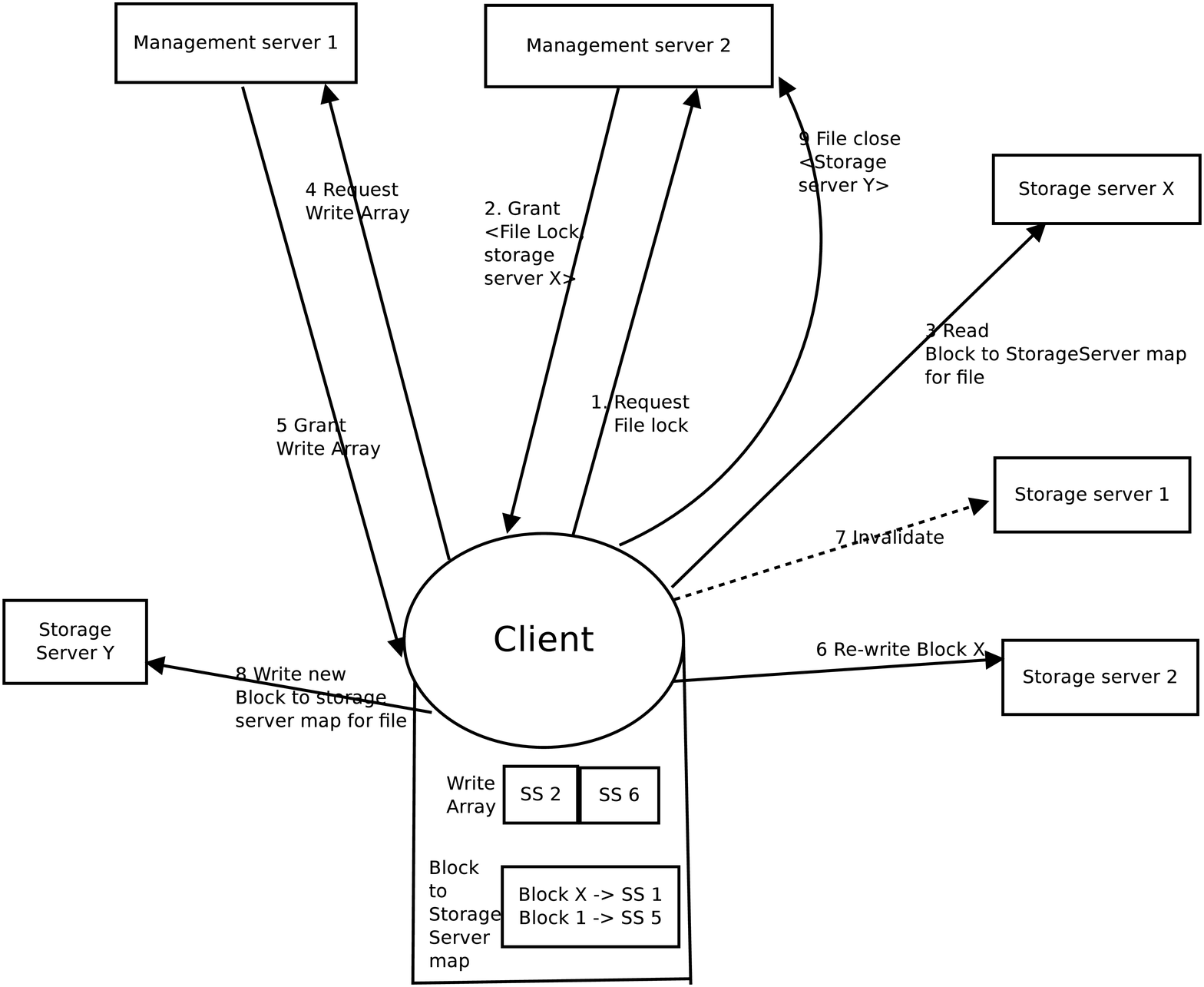}
\caption{Design for distributed wearlevelling}
\label{fig:filelkflow}
%\end{minipage}
\end{figure*}

For wearlevelling, we have to co-ordinate the amount of writes on the storage nodes. Here we present a 
 system model that follows a distributed  co-ordination mechanism that also considers the amount of message communication cost.

The assumptions for our design are as follows:
\begin{itemize}
\item  We assume that our system is made of more than one management servers, many clients and many storage servers.
 
 \item When a file is opened for write, the client request for a file lock. Like in HDFS, we make the
 assumption that there are no concurrent writers. We assume session semantics  where the file write by
 a client is not visible to others until the file is closed.
 \item We do not consider object based devices at storage servers as flash based devices are not sophisticated
 as smart disks. But we assume that there is some local wearlevelling algorithm within a storage node, that makes 
the wear-out of blocks within  a flash storage node even.
 \item We do out-of-place block updates across the storage nodes. When a block is over-written, it is written to a
 different storage server. For this reason, we do not use hashing strategies like Ceph, but use block allocation tables
 to store block to storage server mappings. But the block allocation tables are not maintained by the management server.
 It is situated in one of the storage servers and the management server remembers only the location of the storage server.
 So the metadata that is stored for a single file is less.  This can facilitate the features like \textit{dynamic subtree partitioning}
 of ceph, which can minimise hot spots on a management server.
 \item We have considered all the storage nodes to belong to a single failure domain. So the assumed topology is linear and not a
 hierarchical tree topology.
\end{itemize}

In a disk based storage system, a re-write of a block is targeted to the same disk, where it was previously located. But with 
flash memories, we have to target the re-write to a different storage node for distributing the wear-out on all devices. The storage node to be
 written to should be based on the total `remaining write capacity' of the storage nodes so that nodes are 
worn equally. This requires that we budget the amount of writes to each storage node. Further, in a disk based 
storage system, the decision as to which disk a block has to be written is done by the management server. With flash based distributed 
system, a workload with a lot of re-writes, will involve frequent communication to the management node for deciding the storage node to
 write. 

  We adopted a solution which is followed for distributed quota allocation \cite{digicash}.
  Filesystem quotas restrict the amount of data that users can store. For a distributed filesystem, the  management server has to co-ordinate the quotas for the distributed users 
  operating from the different clients. It is inefficient to involve the management server every time a client has to allocate storage space. 
  The distributed quota allocation follows a digital cash based solution where  the clients receive vouchers from the management servers.
  The number of vouchers denote the quota amount for a client. The clients spend these vouchers  on storage servers for allocating storage space. 
  In this way the management servers are not involved every time when a client allocates space.

  We used a similar approach to the distributed wearlevelling problem.
 With this approach the decision as to which 
flash disk to write is done partly by the management node and partly by the client. The client contacts the management server only periodically 
and not on every write or re-write.

The read/write flow is  illustrated in Figure \ref{fig:filelkflow} and is described below.
\begin{enumerate}
 \item A file id is mapped to management server.  The client which start the I/O operation for a file will request the mapped management
 server for read or write lock.
 \item If the lock has not been granted to some other client, the management server  grants the write lock to the client.  If the file is
 already present, it also gives the storage server X, where the `blocks to storage server mapping' of the file is located.
 \item The client reads from the storage server X, the `blocks to storage server map' for the opened file and stores the information locally.
 \item For doing writes on the file, the client has to have a  ``write-array'' which is a suggestive list of storage servers  and the amount that
 the client can write to  this list of storage servers.   The write array is requested from any management node periodically. The clients again
 request for the write-array if they have run out of the previous one.
 \item The management node fills the requesting client's write-array, with storage servers picked in round robin fashion. The write amounts
 corresponding to each storage server are in proportional to the storage server's remaining write capacity.
 \item Once the client has obtained the write array, the clients use the suggested list of storage servers to do  their out of place 
block level writes without any communication with the management server. For each I/O request, the client  picks the first storage server
 in round robin fashion, from the write-array, which has enough write amount left. The client targets the block write to the picked storage 
server and also updates the local copy of the `blocks to storage server map'
  \item In case of re-write of a block, the client has to invalidate the old block on the old storage server location.
 \item Before closing the file, the client writes the new `blocks to storage server map' to another storage server Y.
 \item When the file is closed finally, the client intimates the management server of the new location of the `blocks to storage server map'.
  The management server withdraws the lock it has granted and also updates its tables about the new map-file location. 
\end{enumerate}

 The management server has to know the approximate wear and space utilisation information of the storage servers.
 This information has to be obtained, every once in a while, by one of the management servers by querying the storage servers. 
 This is similar to the reconciliation phase in \cite{digicash}.
 A threshold ``$\phi$'' defines the maximum total write amount, a management server can suggest to clients, before reconciliation. 
A management server which has suggested some ``$\phi$'' write  amount to the clients, takes on the responsibility  to request the storage servers for
 their remaining write capacity and subsequently synchronises this information across other management nodes.

\subsubsection{Delayed Invalidations}
  As an optimisation, it is possible for the clients to delay the invalidates and batch several invalidates to the storage servers.
 Longer delays can result in lesser message communication. But delaying too much will cause the remaining space  to run out on the storage servers.
 So to overcome this problem, we delayed the invalidate messages depending upon the free space on a storage server.

    With this approach, along with the ``write-array'' the management server gives the last known space utilisation of the storage servers. 
So the clients keep track of  the approximate space utilisation of the storage servers. Depending on the amount of free space available, the clients
 delay the amount of invalidation messages to the storage servers. We maintain different delay thresholds for different space utilisations. 
When the number of invalidation messages that are delayed cross the threshold, the invalidates are batched together and sent to the storage server.

\section{Simulation Results}

\ifdefined\THESIS

\else
\begin{figure*}[t]

\begin{minipage}[b]{0.5\columnwidth}
 \centering
\end{minipage}

\hspace{1.8cm}
\begin{minipage}[b]{0.5\columnwidth}
 \centering
\includegraphics[width=2.5in,height=1.5in]{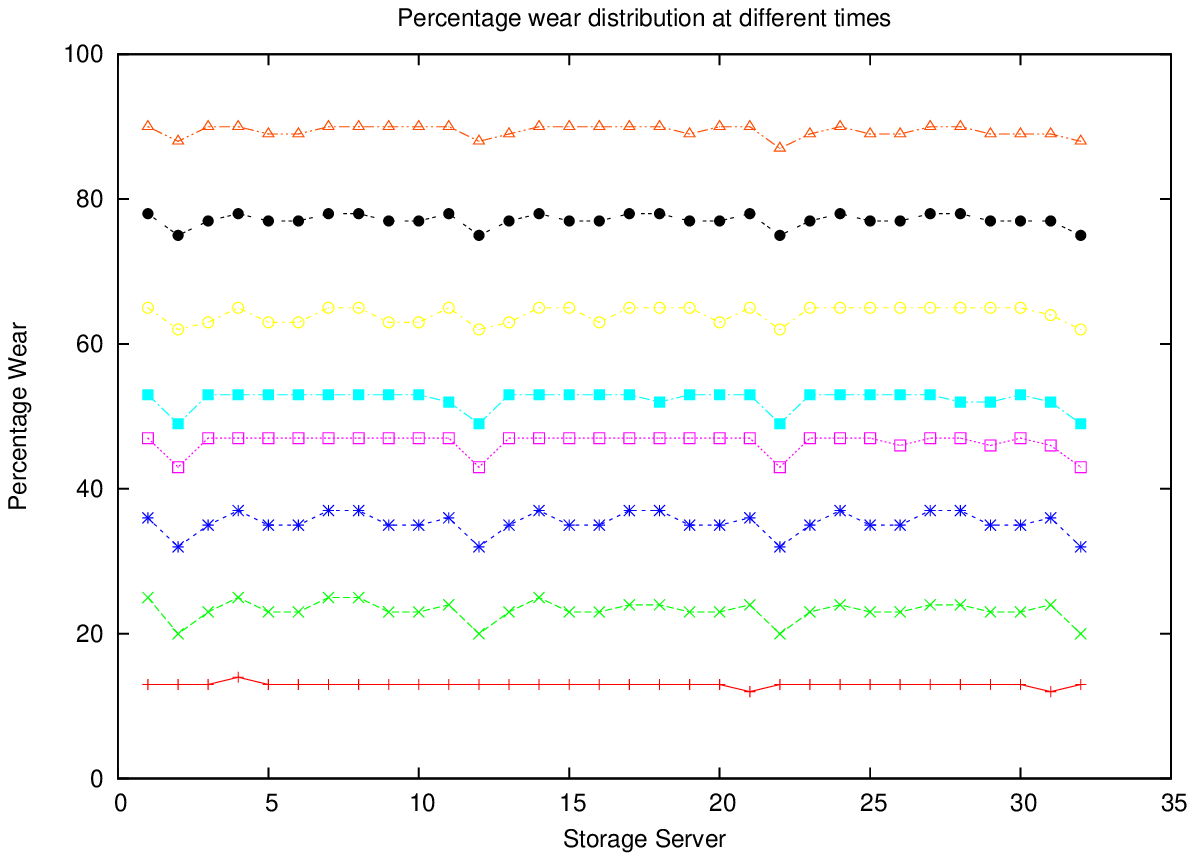}
\caption{Percentage wear}
\label{fig:wearpercent}
\end{minipage}
\hspace{1.8cm}
\begin{minipage}[b]{0.5\columnwidth}
\includegraphics[width=2.5in,height=1.5in]{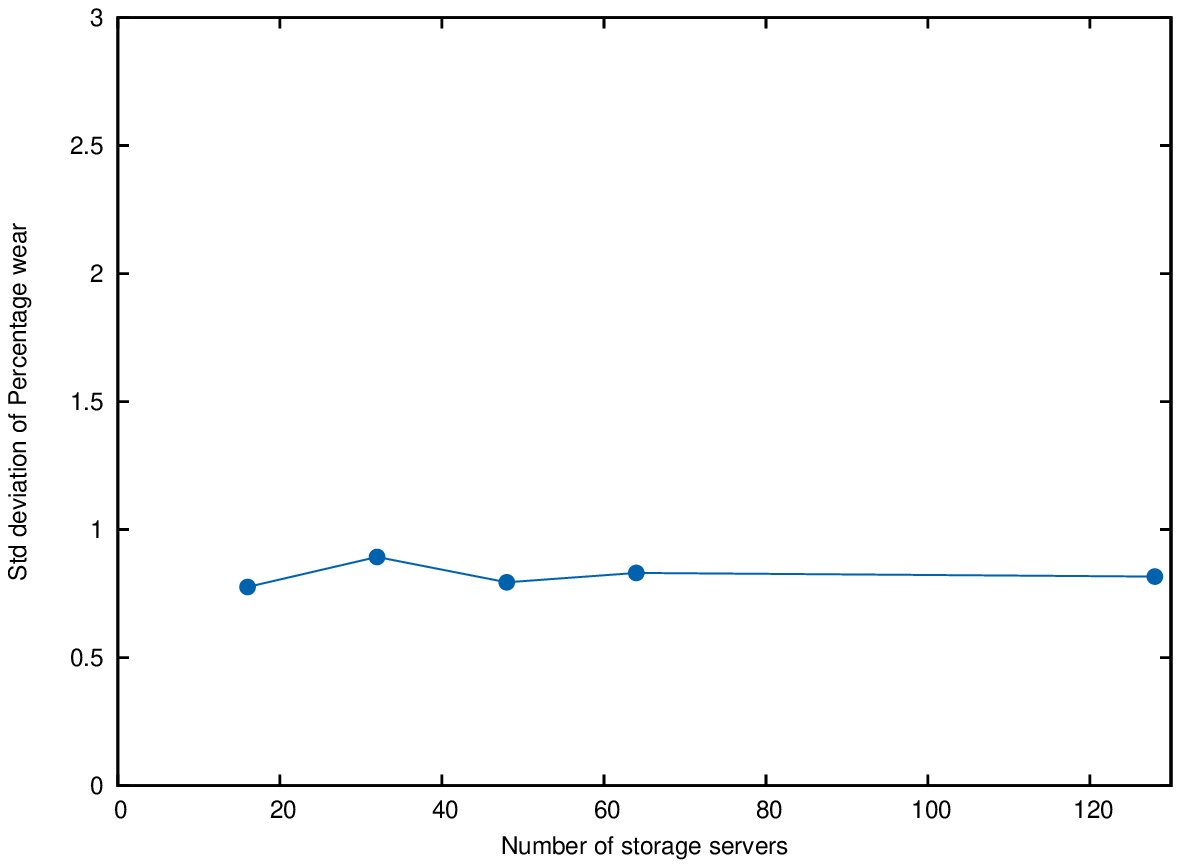}
\caption{Standard deviation of wear percent}
\label{fig:stddev}
\end{minipage}

\end{figure*}
\fi

We developed a discrete event based simulator for a distributed storage system consisting of  management servers, storage servers and clients. 
 Events generated from clients, management servers and storage servers are floated in the event queue. A  network latency of 250us is added
 to the events that are targeted to other nodes. Each of the clients, storage server and management server has an associated queue.

  At the storage server end, we emulate the  a nand block device as a contiguous array with each element in the array storing the statistical information 
for the blocks. We  implements nand block read, write and invalidates as modifications to the emulated nand block array. The flash page read, page 
write and block erase latencies are 25us, 200us, 2000us respectively. The I/O request latency will be equal to queuing latency plus  the flash device
 read or write latencies. We did not add the garbage collection overhead to the latencies.

  We assumed a workload with  vast number of accesses to small files. This access pattern is common for distributed workloads\cite{measuredistfs}.
 So our workload consist of writes and re-writes to a number of small files. Each file has upto 5-8 blocks. With a distributed filesystem block 
size as 64MB, the maximum size of a file created is 512MB.  
 
  We assumed a heterogeneous storage system made of flash storage nodes with different capacities and different maximum endurance cycles.
 We varied  the maximum endurance cycles  between 100 to 400. The plot in Figure \ref{fig:wrcap} shows the distribution of the initial 
write capacity (Number of blocks in storage node * Maximum endurance cycles) of each storage node.  The number of storage servers is 32 in our simulation. 

\ifdefined\THESIS

\begin{figure}[htb]
  \vspace{9pt}

  {\hbox{ \hspace{1.5in} 
    \epsfxsize=3.4in
    \epsffile{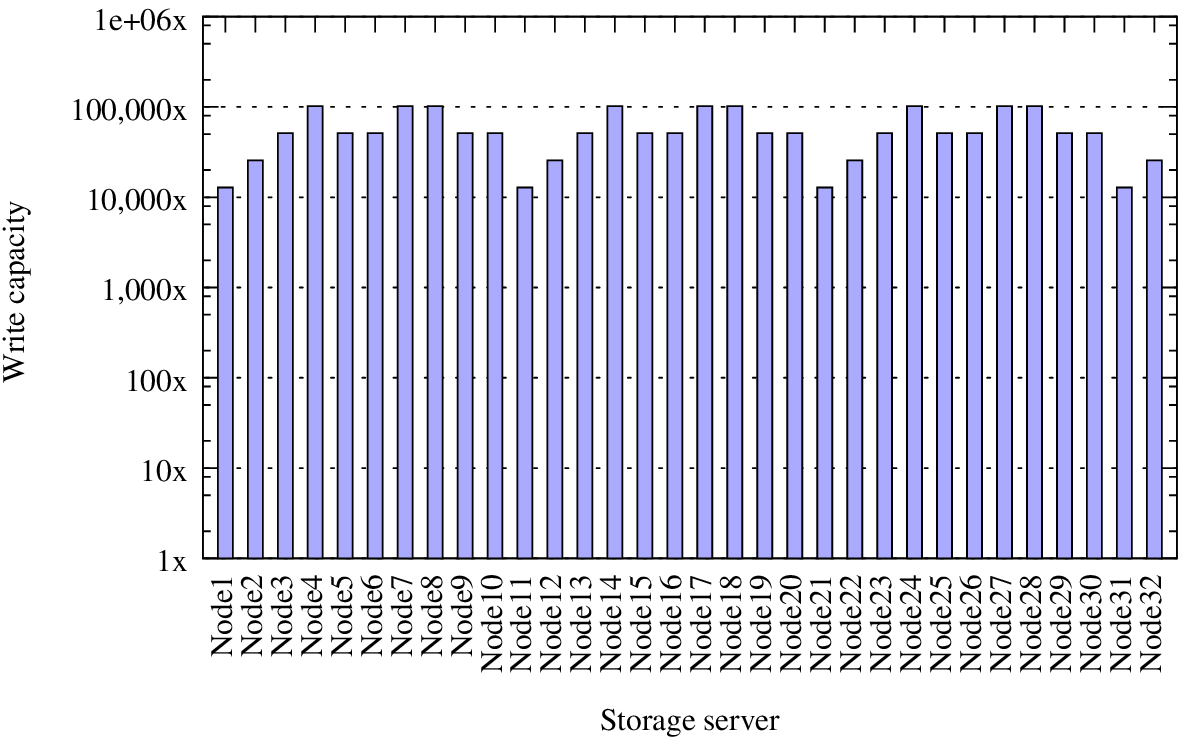}
    \hspace{0.25in}
    
    }
  }

  \caption{Initial Write capacities: Write capacity is the maximum  amount of data that can be written to the storage node in its lifetime. Write capacity is plotted in the unit  of  blocks.}
  \label{fig:wrcap}

\end{figure}

\FloatBarrier

\else
\begin{figure}[h]

\begin{minipage}[b]{0.5\columnwidth}
\includegraphics[width=3.0in,height=1.5in]{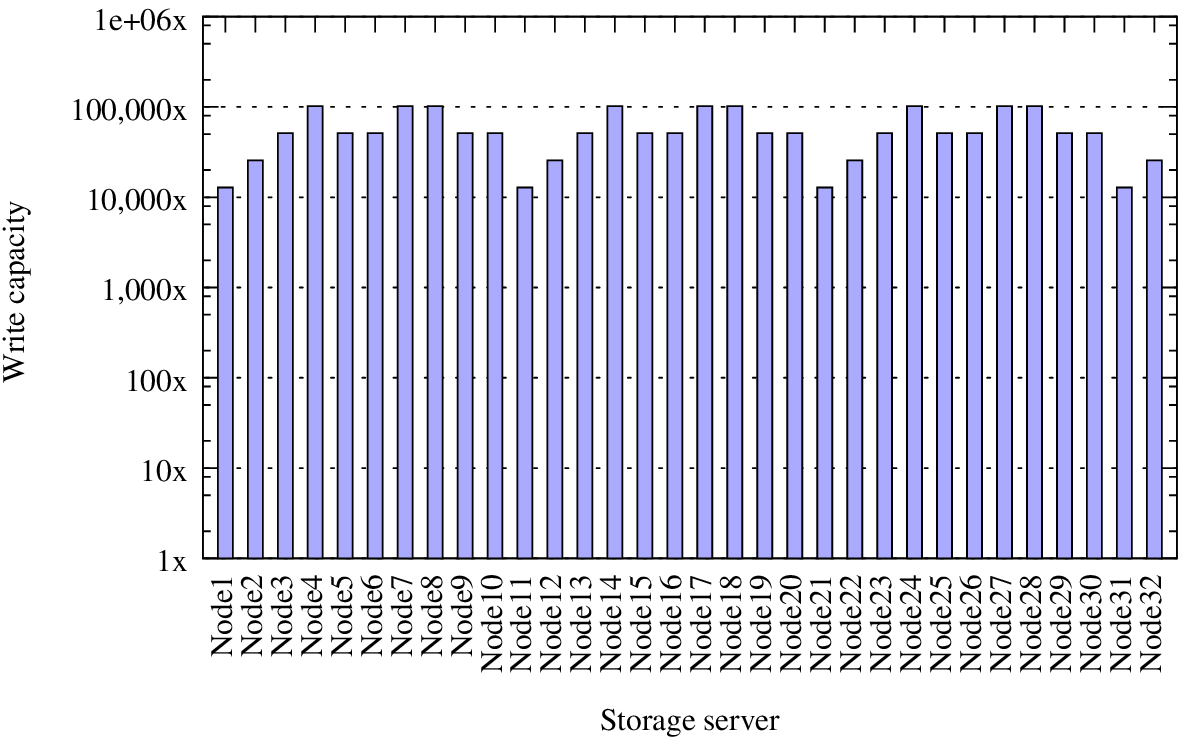}
\caption{Initial Write capacities}
\label{fig:wrcap}
\end{minipage}
\end{figure}
\fi

  Assuming an ideal local wearlevelling algorithm within a flash storage node, we calculated a metric  ``\textit{percentage-wear}'' for a node as
 ``((Total amount of bytes written so far) *100)/(Storage node size*maximum erase cycles)''. 

  Only write I/O requests are input to the simulation setup for an initial period of time, to allow the warming up of the tables 
and data structures.  After the initial warm up time the results are taken.

 The distribution of Percentage-wear of flash disks at different instances of  the simulation is shown in Figure \ref{fig:wearpercent}. All the storage
 nodes age in the same manner. The plot in Figure \ref{fig:stddev} show the standard deviation of percentage-wear at the end of simulation for different 
number of storage nodes. We were able to achieve standard deviation values to be less than one. 

\ifdefined\THESIS

\begin{figure}[htb]
  \vspace{9pt}

  \centerline{\hbox{ \hspace{0.0in} 
    \epsfxsize=3.4in
    \epsffile{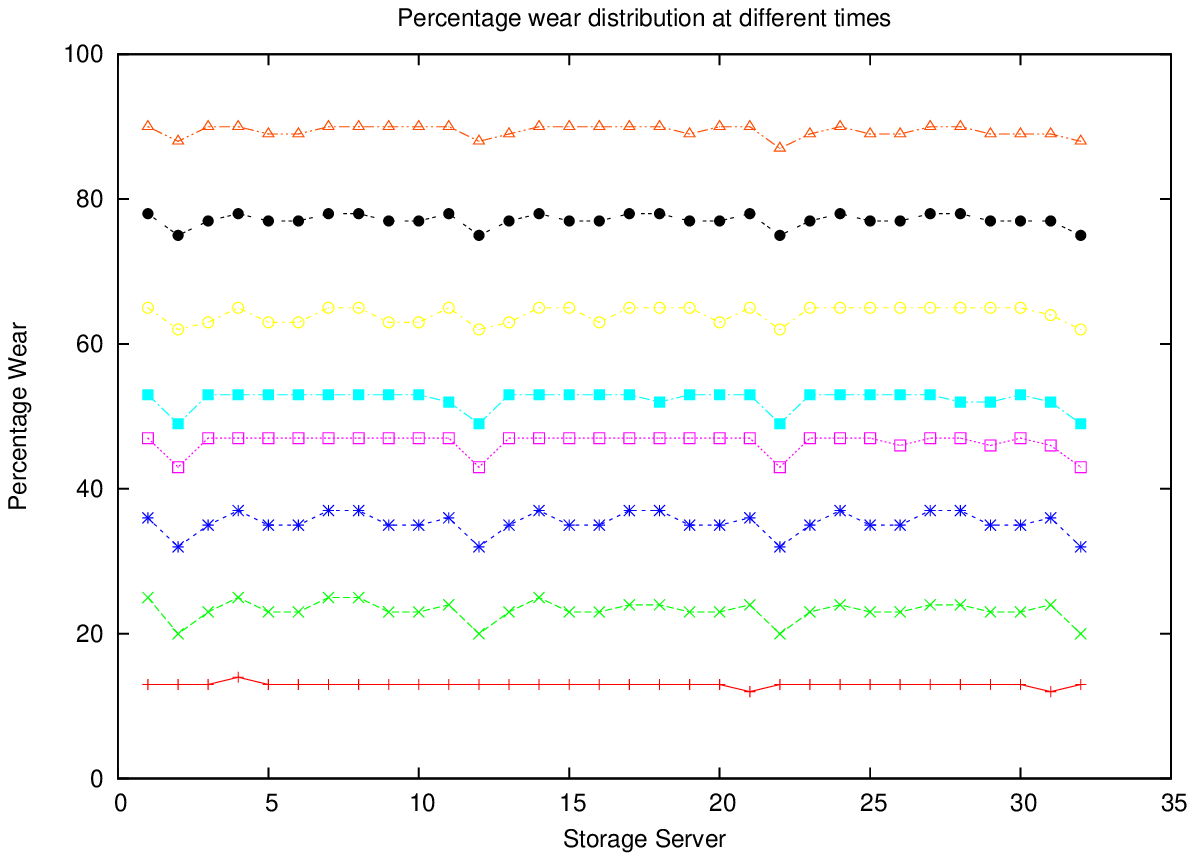}
    \hspace{0.25in}
    \epsfxsize=3.4in
    \epsffile{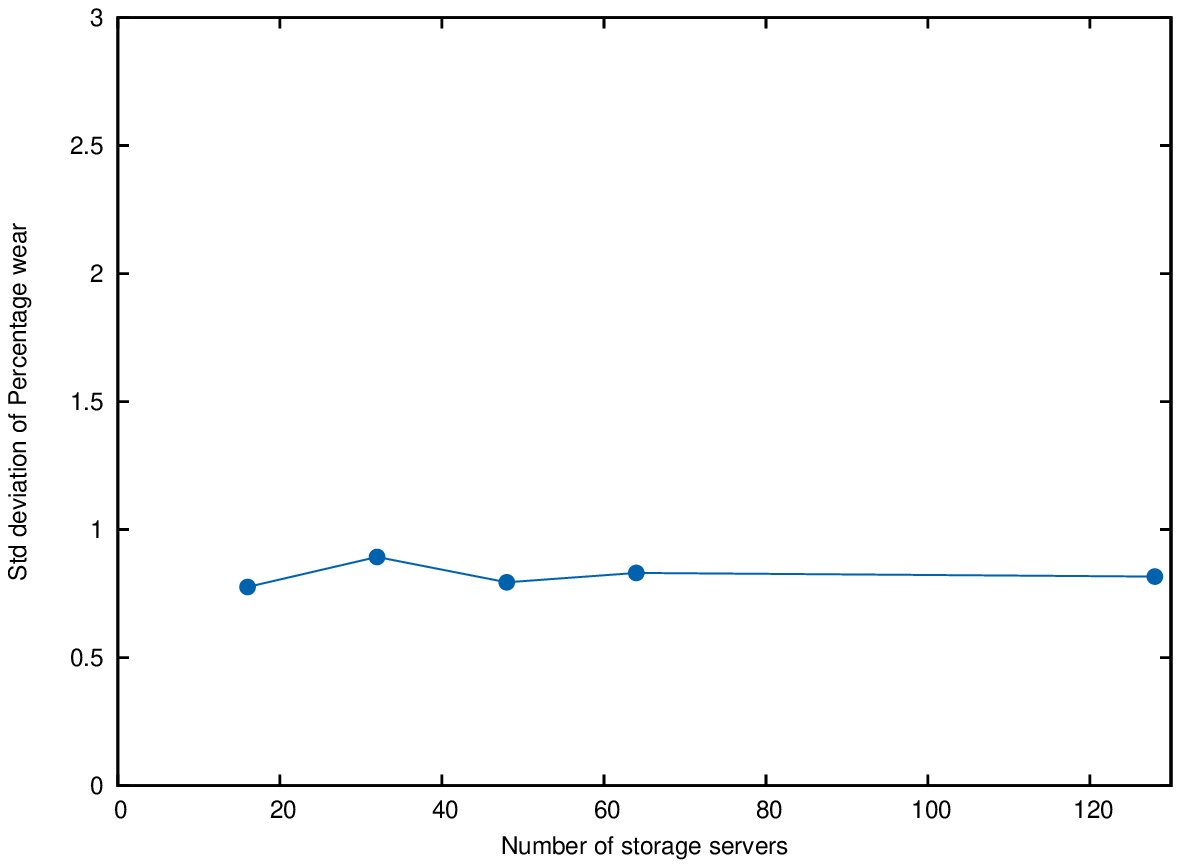}
    }
  }

  \caption{ This Figure contains two plots (a)Percentage wear of the storage nodes at different instants of time during the course of simulation.  (b) Final Standard deviation of percentage-wear vs storage system size.}
  \label{fig:wearsd}

\end{figure}

\else

\fi

\begin{table*}[]
\centering

    \caption{Number of invalidation related messages}     % NOTE!  caption goes _before_ the table contents !!
    \label{tab:invmesg}

    \begin{small}	
    \begin{tabular}{|l|l|l||}
    \hline
    {\bfseries Number of} &  {\bfseries  Without Delayed }         & {\bfseries Delayed }    \\
    
    {\bfseries Storage nodes} & {\bfseries  Invalidates}         & {\bfseries  invalidates}    \\
    \hline
32 & 1,662,355	&	183,395  	\\
\hline
    \end{tabular}
    \end{small} 
\end{table*}

The table \ref{tab:invmesg} shows the number of invalidation related messages, in a 32 node storage system. The number of messages when
 invalidates are not batched together is above a million. For the delayed invalidates case, we batch together several invalidation messages
 depending on the space utilisation. When the space utilisation of a storage server is above 50\%, we don't delay the invalidates.
 We vary the number of delayed invalidation messages from 7 to 15 as the space utilisation varies from less than 50\% to less than 20\%

\ifdefined\THESIS
\FloatBarrier
\fi
\begin{comment}

\begin{table*}[]
\centering

    \caption{Number of invalidation related messages}     % NOTE!  caption goes _before_ the table contents !!
    \label{tab:invmesg}

    \begin{small}	
    \begin{tabular}{|l|l|l|l|}
    \hline
    {\bfseries Number of} & \multicolumn{3} {c|} {\bfseries number of invalidate related messages} \\
    \cline{2-4}
    {\bfseries Storage nodes} & {\bfseries  No Delayed Inv}         & {\bfseries Cli trig GC}     & {\bfseries StgServ trig GC}           \\
    \hline
    12         & 619,162	&	40,210 	& 41,549 	\\
              & 		&		& 		\\

    \hline
    22         & 1,206,307 & 78,746 & 58,943	\\
              &			&		&    	\\
              
    \hline
    32        & 1,799,609 	&	117,755	& 77,002      \\
              & 	&		& 	\\
              
    \hline
    42        &	2,385,803 	&	156,170 	&	137,435 \\
    \hline
    
    \hline
    \end{tabular}
    \end{small} 
\end{table*}
The table \ref{tab:invmesg} shows the number of invalidation related messages for the case when there is no delayed garbage collection, 
when the delayed garbage collection is triggered by the client and storage server. We see that though the storage server requests all the
 clients for invalidates, storage server triggered garbage collection incurs lesser number of messages.  
\end{comment}
% conference papers do not normally have an appendix

\section{Related work}
\begin{comment}
Dave roberts thesis
Digital cash Quota allocation
RAID
\end{comment}

Wearlevelling across flash devices is mostly considered in the case of flash arrays.  Particularly the usage of flash memories for RAID arrays
 is discussed in \cite{diffRAID}\cite{Erasurecoding}.  In \cite{diffRAID}, they take in to account, the fact that, bit error rate of a drive 
increases as it grows older. So they have  proposed, to use uneven parity assignment and wear out a older drive faster, in order that they 
can be replaced early. This early replacement of an error-prone drive improves the reliability of the entire RAID array. In \cite{Erasurecoding} they
 have tried to balance the load across the flash array by making use of erasure coding techniques. The NAND chip array is divided into zones and the data of most
 frequently accessed zones are coded and stored in another flash chip. When the data of heavily loaded chip needs to be accessed, instead of doing the I/O directly 
from the heavily loaded chip, they reconstruct the data using coding techniques.

  Compared to the these approaches which take care of the flash memory wear leveling issues for RAID like scenarios, we have considered the flash wear
 leveling problem in the distributed file system level where the storage nodes are more dispersed by network than in a storage array.  For the network distributed
 storage nodes, other metrics like reducing the message communication cost is also vital.

 To the best of our knowledge, the thesis in reference \cite{DaveRoberts} is  the only previous work with respect to wearlevelling with flash devices spread across a network. 
The thesis \cite{DaveRoberts}  only considers migration of data across SSDs in order to evenise the wear.  They have not considered the problem from the data distribution perspective as our work does.   

    The use of a `suggestive write-array' in our design is inspired by the digital cash based algorithm used in distributed quota
 allocation in \cite{digicash}. In distributed quota allocation problem we have to limit the amount written by the client. The distributed wear
 leveling problem is the symmetrical case where we have to limit the amount written to the storage server.

\section{ Future work and Conclusion}
 
 In a distributed storage system, data migration happens because of addition of new storage in a disk or removal of storage.
 Using these migration mechanisms to move cold (not frequently modified) data around will be an interesting approach, and we
 are exploring this for future work.  Currently,  we have not considered placement across failure domains, based on topology.
 Spreading the replicas across failure domains, ensuring distributed wearlevelling is another interesting direction for future work. 
 For our simulation, we have assumed a same idealized wear leveling algorithm on all the storage servers. Implementing 
 different local wear leveling algorithms on different nodes and evaluate them for any interference will be one more future direction of work.

In conclusion, we have addressed the wear leveling problem for distributed flash memories from the distributed file system perspective.
 We have motivated the problem, by first considering the HDFS replica placement.  We have then suggested a design and an algorithm to do wearlevelling across
 the flash nodes and also minimises communication cost between client and management servers.  We have also  suggested a method for delaying garbage collection to
 further minimise the  number of messages.

 \begin{comment}
% use section* for acknowledgement
\section*{Acknowledgment}

The authors would like to thank...
more thanks here
\end{comment}

{\bibliographystyle{abbrv}
  \bibliography{dist}
}

\begin{comment}
 
\end{comment}

% that's all folks
\end{document}